\documentclass[a4paper,twoside]{article}
\pdfoutput=1 

\usepackage{epsfig}
\usepackage{subcaption}
\usepackage{calc}
\usepackage{amssymb}
\usepackage{amstext}
\usepackage{amsmath}
\usepackage{amsthm}
\usepackage{bm}
\usepackage{multicol}
\usepackage{pslatex}
\usepackage{apalike}
\usepackage{algorithm2e}
\usepackage{algpseudocode}
\usepackage[bottom]{footmisc}
\usepackage{braket}
\usepackage{url}
\usepackage{SCITEPRESS}     % Please add other packages that you may need BEFORE the SCITEPRESS.sty package.

\begin{document}

\title{Embedding of Tree Tensor Networks into Shallow Quantum Circuits}

% \author{\authorname{Shota Sugawara\sup{1}\orcidAuthor{0009-0005-8758-2940},Kazuki Inomata\sup{1}\orcidAuthor{0009-0008-4030-5518}, Tsuyoshi Okubo\sup{2}\orcidAuthor{0000-0003-4334-7293} and Synge Todo\sup{1,2,3}\orcidAuthor{0000-0001-9338-0548}}
% \affiliation{\sup{1}Department of Physics, The University of Tokyo, Tokyo, 113-0033, Japan}
% \affiliation{\sup{2}Institute for Physics of Intelligence, The University of Tokyo, Tokyo, 113-0033, Japan}
% \affiliation{\sup{3}Institute for Solid State Physics, The University of Tokyo, Kashiwa, 277-8581, Japan}
% \email{\{shota.sugawara, kazuki.inomata, t-okubo and wistaria\}@phys.s.u-tokyo.ac.jp}
% }

\author{\authorname{Shota Sugawara\sup{1} ,Kazuki Inomata\sup{1}, Tsuyoshi Okubo\sup{2} and Synge Todo\sup{1,2,3}}
\affiliation{\sup{1}Department of Physics, The University of Tokyo, Tokyo, 113-0033, Japan}
\affiliation{\sup{2}Institute for Physics of Intelligence, The University of Tokyo, Tokyo, 113-0033, Japan}
\affiliation{\sup{3}Institute for Solid State Physics, The University of Tokyo, Kashiwa, 277-8581, Japan}
\email{\{shota.sugawara, kazuki.inomata, t-okubo and wistaria\}@phys.s.u-tokyo.ac.jp}
}


\abstract{Variational Quantum Algorithms~(VQAs) are being highlighted as key quantum algorithms for demonstrating quantum advantage on Noisy Intermediate-Scale Quantum~(NISQ) devices, which are limited to executing shallow quantum circuits because of noise.
However, the barren plateau problem, where the gradient of the loss function becomes exponentially small with system size, hinders this goal.
Recent studies suggest that embedding tensor networks into quantum circuits and initializing the parameters can avoid the barren plateau.
Yet, embedding tensor networks into quantum circuits is generally difficult, and methods have been limited to the simplest structure, Matrix Product States~(MPSs).
This study proposes a method to embed Tree Tensor Networks~(TTNs), characterized by their hierarchical structure, into shallow quantum circuits.
TTNs are suitable for representing two-dimensional systems and systems with long-range correlations, which MPSs are inadequate for representing.
Our numerical results show that embedding TTNs provides better initial quantum circuits than MPS.
Additionally, our method has a practical computational complexity, making it applicable to a wide range of TTNs.
This study is expected to extend the application of VQAs to two-dimensional systems and those with long-range correlations, which have been challenging to utilize.}

\onecolumn \maketitle \normalsize \setcounter{footnote}{0} \vfill

\section{\uppercase{Introduction}}
Variational Quantum Algorithms~(VQAs) represent the foremost approach for achieving quantum advantage with the current generation of quantum computing technologies.
Quantum computers are anticipated to outperform classical ones, with some algorithms already proving more efficient~\cite{shor1994algorithms,grover1996fast}.
However, the currently available Noisy Intermediate-Scale Quantum~(NISQ) devices are incapable of executing most algorithms due to their limited number of qubits and susceptibility to noise~\cite{nisq}.

VQAs are hybrid methods where classical computers optimize parameters of quantum circuits' ansatz to minimize the cost function evaluated by quantum computers.
VQAs require only shallow circuits, making them notable as algorithms that can be executed on NISQ devices~\cite{cerezo2021variational}.
VQAs come in many forms, such as Quantum Machine Learning~(QML)~\cite{qml0,qml1,qml2}, Variational Quantum Eigensolver~(VQE)~\cite{vqe0,vqe1,vqe2}, and Quantum Approximate Optimization Algorithm~(QAOA)~\cite{qaoa0,qaoa1,qaoa2}, with potential applications across diverse industries and fields.

However, a significant challenge known as the barren plateau stands in the way of realizing quantum advantage~\cite{barrenplateau1,barrenplateau2}.
The barren plateau phenomenon refers to the challenge in VQAs where the gradient of the cost function decreases exponentially as the system size increases.
This phenomenon occurs regardless of whether the optimization method is gradient-based~\cite{gradient-based} or gradient-free~\cite{gradient-free}, and it has been observed even in shallow quantum circuits~\cite{barrenplateau-shallow}.
Furthermore, it has been confirmed that this phenomenon also occurs in practical tasks using real-world data~\cite{barren-generic0,barren-generic1,barren-generic2}.
Avoiding the barren plateau is a critical challenge in demonstrating the superiority of quantum algorithms using NISQ devices.
To avoid the barren plateau, appropriate parameter initialization in VQAs is crucial since randomly initializing the parameters can result in the algorithm starting far from the solution or near a local minimum~\cite{qaoa2}.
Although various initialization methods have been considered~\cite{avoid-barren0,avoid-barren1,avoid-barren2}, using tensor networks is natural due to their compatibility with quantum circuits.

Tensor networks are originally developed to efficiently represent quantum many-body wave functions.
Any quantum circuit can be naturally regarded as a tensor network~\cite{tn-qc}, and it is sometimes possible to simulate quantum computers with a practical amount of time using tensor networks~\cite{google-ibm}.
Moreover, in recent years, their utility has been recognized and applied across various fields such as machine learning~\cite{tnml} and language models~\cite{tnlanguage}

In this study, we focus particularly on Matrix Product States~(MPSs) and Tree Tensor Networks~(TTNs) among various structures of tensor networks.
MPSs are one-dimensional arrays of tensors.
Its simplest and easiest-to-use structure, along with the presence of advanced algorithms such as Density Matrix Renormalization Group~(DMRG)~\cite{dmrg} and Time-evolving block decimation~(TEBD)~\cite{tebd}, has led to its application across a wide range of fields.
Recently, these excellent algorithms have been applied to the field of machine learning, continuing the exploration of new possibilities~\cite{MPSsupervised,MPSgenerative}.
TTNs are tree-like structures of tensors.
The procedures used in DMRG and TEBD have been applied to TTNs~\cite{dmrg-ttn,tebd-ttn}.
Additionally, the distance between any pair of leaf nodes scales in logarithmic order in TTNs while in linear order in MPSs.
As connected correlation functions generally decay exponentially with path length within a tensor network, TTNs are better suited to capture longer-range correlations and represent two-dimensional objects than MPSs.
TTNs are utilized in various fields such as chemical problems~\cite{chemical1,chemical2}, condensed matter physics~\cite{condmat1,condmat2,condmat3} and machine learning~\cite{ml1,ml2}.

Rudolph et al. have successfully avoided the barren plateau problem by utilizing tensor networks~\cite{rudolph2023synergistic}.
They proposed a method of first optimizing tensor networks, then mapping the optimized tensor networks to quantum circuits, and finally executing VQAs.

\begin{figure}
    \centering
    \includegraphics[width=\linewidth]{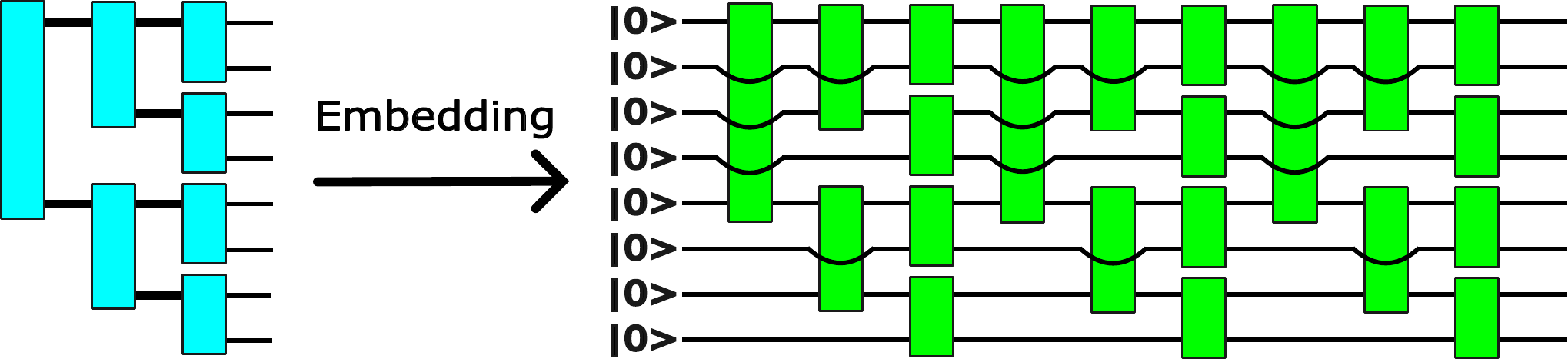}
    \caption{A schematic diagram illustrating the embedding of a TTN into a shallow quantum circuit composed solely of two-qubit gates. The aim is to ensure that the quantum state output by the quantum circuit closely approximates the quantum state represented by the TTN. The diagram depicts a case with three layers, and the approximation accuracy improves as the number of quantum circuit layers increases.}
    \label{fig:task-setting}
\end{figure}
% \begin{figure}[!h]
%   \centering
%    {\epsfig{file = task-setting.pdf, width = 7cm}}
%   \caption{A schematic diagram illustrating the embedding of a TTN into a shallow quantum circuit composed solely of two-qubit gates. The aim is to ensure that the quantum state output by the quantum circuit closely approximates the quantum state represented by the TTN. The diagram depicts a case with three layers, and the approximation accuracy improves as the number of quantum circuit layers increases.}
%   \label{fig:task-setting}
%  \end{figure}
While numerical results have shown that this method can indeed avoid the barren plateau, there is a general challenge in mapping tensor network states into shallow quantum circuits.
The case where the tensor network structure is an MPS has already been well-studied~\cite{EncodingMPS,mpsdecomp,mpsoptim,mpspreparation}.
However, effective methods for mapping tensor networks other than MPSs to shallow quantum circuits have not yet been devised.

In this paper, we propose a method to embed TTNs into shallow quantum circuits composed of only two-qubit gates as shown in Figure~\ref{fig:task-setting}.
The primary obstacle to embedding TTNs into shallow quantum circuits has been the complexity of contractions arising from the intricate structure of TTNs.
In general, the contraction of tensor networks requires an exponentially large memory footprint relative to the number of qubits if performed naively.
Furthermore, determining the optimal contraction order for tensor networks is an NP-complete problem~\cite{np-complete}.
Through the innovative design of a contraction method that balances minimal approximation error and computational efficiency, we successfully extend the embedding method of MPSs~\cite{mpsdecomp} to TTNs.
Additionally, by applying the proposed method to practical problems, we successfully prepare better initial quantum circuits for VQAs than those provided by MPSs with a practical computational complexity.
This study is expected to extend the application of VQAs to two-dimensional systems and those with long-range correlations, which have previously been challenging to utilize.

\section{\uppercase{Background}}
\subsection{Tensor Networks}
Tensor networks are powerful mathematical frameworks that efficiently represent and manipulate high-dimensional data by decomposing tensors into interconnected lower-dimensional components~\cite{tn-review,tn-review2}.
They have been widely applied in various fields, including quantum many-body physics for approximating ground states of complex systems~\cite{dmrg,tebd,tn-review3} and machine learning for real data~\cite{MPSsupervised,MPSgenerative}.
Recently, increasing attention has been directed towards their compatibility with quantum computing, as their structure aligns well with quantum circuits and facilitates hybrid quantum-classical algorithms~\cite{tn-qc}.

The widespread application of tensor networks across various fields can be attributed to their simple and comprehensible notation.
Tensor network notation provides a clear and intuitive way to represent complex tensor contractions and operations.
In tensor network notation, a rank-$r$ tensor is depicted as a geometric shape with $r$ legs.
The geometric shape and the direction of the legs are determined by the properties of the tensor and its indices.
When representing quantum states, the direction of the legs often indicates whether the vectors are in the Hilbert space for kets or its dual space.
When two tensors share a single leg, the leg is referred to as the bond, and the dimension of that leg is referred to as the bond dimension. 
By adjusting the bond dimension, the expressive power of the tensor network can be controlled. 
Reducing the bond dimension to decrease memory requirements is referred to as truncation. 
One of the most commonly used operations is contraction, which combines multiple tensors into a single tensor. 
In tensor network diagrams, contraction corresponds to connecting tensors with lines.
The contraction of $A$'s $x$-th index and $B$'s $y$-th index is defined as
\begin{equation}
\begin{aligned}
    C_{i_1,\dots,i_{x-1},i_{x+1},\dots,i_r,j_1,j_{y-1},j_{y+1},\dots,j_s} \\
    =\sum_\alpha A_{i_1,\dots,i_{x-1},\alpha,i_{x+1},i_r}B_{j_1,\dots,j_{y-1},\alpha,j_{y+1},\dots,j_s},
\end{aligned}
\end{equation}
where $C$ is the result of the contraction.
Operations such as the inner product of vectors, matrix-vector multiplications, and matrix-matrix multiplications are specific examples of contractions.
A contraction network forms a tensor network, allowing for the consideration of various networks depending on the objective.

\subsection{MPSs and TTNs}
Let $\ket{\psi}$ be a tensor network in the form of either an MPS or a TTN.
Number the tensors in $\ket{\psi}$ from left to right for MPS, and in breadth-first search~(BFS) order from the root node for TTN, denoting the $i$-th tensor as $A^{(i)} (i=1,\dots, N)$.
By adding a leg with bond dimension one, a two-legged tensor can be converted into a three-legged tensor, thus all tensors in $\ket{\psi}$ can be considered three-legged.
$\ket{\psi}$ is in canonical form if there exists a node $A^{(i)}$, called the canonical center, such that for all other nodes $A^{(i')}$ the following holds
\begin{equation}
    \sum_{l,m}A^{(i')}_{l,m,n}A^{(i')*}_{l,m,n'}=I_{n,n'},
\end{equation}
where the leg denoted by index $n$ is the unique leg of $A^{(i')}$ pointing towards $A^{(i)}$.
A tensor that satisfies this equation is referred to as an isometric tensor.
Any $\ket{\psi}$ can be transformed into this canonical form and the position of the canonical center can be freely moved without changing the quantum state.

The canonical form offers numerous advantages.
In this paper, the key benefit is that at the canonical center, the local Singular Value Decomposition~(SVD) matches the global SVD, allowing for precise bond dimension reduction through truncation while appropriately moving the canonical center.
Additionally, in the canonical form, each tensor is an isometry, facilitating easy conversion to unitary form and embedding into quantum circuits. 
Unless otherwise specified, this paper assumes that any MPS and TTN are converted to the canonical form with $A^{(0)}$ as the canonical center.

\subsection{Embedding of MPSs}
\begin{figure*}
    \centering
    \includegraphics[width=\linewidth]{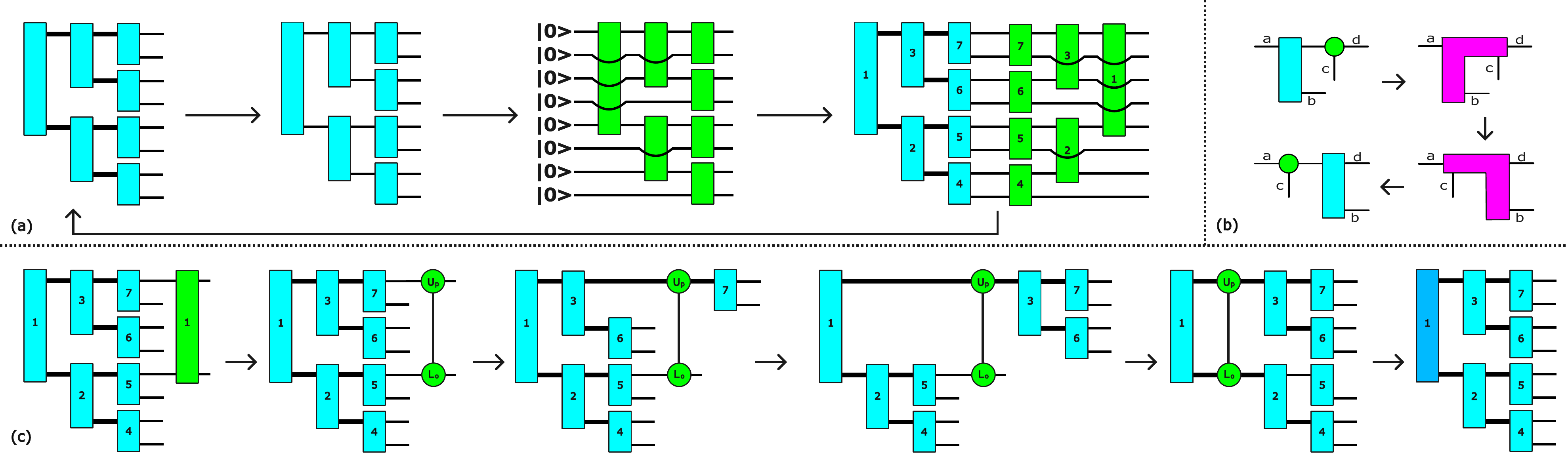}
  \caption{A schematic diagram illustrating the systematic decomposition for a TTN. (a) The TTN is truncated to a bond dimension of two using SVD. This simplified TTN is then embedded into a quantum circuit by converting tensors into unitaries via Gram-Schmidt orthogonalization. These unitaries act as disentanglers and are applied to the original TTN. By repeating these steps, a multi-layer quantum circuit is formed. (b) A diagram of a penetration algorithm. It contracts two connected tensors, reorders axes, and separates them using SVD, making it seem like their positions have swapped. (c) A schematic diagram explaining the transformation from a complex tensor network into a TTN. We number the tensors based on their positions in the network. We first use SVD to split it into an upper and a lower tensor. Then, we apply the penetration algorithm iteratively until the upper tensor is connected with the corresponding tensor. This process is repeated for the lower tensor as well. Finally, we contract the upper and lower tensors with the corresponding tensor to form a new tensor in the TTN. This process is performed sequentially from the highest-numbered component of the disentangler.}
  \label{fig:decomposition}
\end{figure*}
% \begin{figure*}[!h]
%   \centering
%    {\epsfig{file = fig-decomposition.pdf, width = 15cm}}
%   \caption{A schematic diagram illustrating the systematic decomposition for a TTN. (a) The TTN is truncated to a bond dimension of two using SVD. This simplified TTN is then embedded into a quantum circuit by converting tensors into unitaries via Gram-Schmidt orthogonalization. These unitaries act as disentanglers and are applied to the original TTN. By repeating these steps, a multi-layer quantum circuit is formed. (b) A diagram of a penetration algorithm. It contracts two connected tensors, reorders axes, and separates them using SVD, making it seem like their positions have swapped. (c) A schematic diagram explaining the transformation from a complex tensor network into a TTN. We number the tensors based on their positions in the network. We first use SVD to split it into an upper and a lower tensor. Then, we apply the penetration algorithm iteratively until the upper tensor is connected with the corresponding tensor. This process is repeated for the lower tensor as well. Finally, we contract the upper and lower tensors with the corresponding tensor to form a new tensor in the TTN. This process is performed sequentially from the highest-numbered component of the disentangler.}
%   \label{fig:decomposition}
%  \end{figure*}
 
Although embedding general tensor networks into shallow quantum circuits is challenging, several workable methods have been proposed for MPSs.
In this subsection, we overview the technique for seamlessly embedding MPSs into shallow quantum circuits.

% systematical decomposition by disentangling
Ran introduced a systematic decomposition method for an MPS into several layers of two-qubit gates with a linear next-neighbor topology~\cite{EncodingMPS}. 
First, the MPS $\ket{\psi^{(k)}}$ with bond dimension $\chi$ is truncated to a bond dimension two MPS $\ket{\psi_{\chi=2}^{(k)}}$.
Next, the isometric tensors in $\ket{\psi_{\chi=2}^{(k)}}$ are converted into unitary tensors.
The resulting set of unitary tensors, $L[U]^{(k)}$, is referred to as a layer and can be embedded into a quantum circuit composed of two-qubit gates.
Additionally, since
\begin{equation}
    \ket{\psi_{\chi=2}^{(k)}}=\mathrm{L}[U]^{(k)}\ket{0}
\end{equation}
and
\begin{equation}
    \mathrm{L}[U]^{(k)\dagger}\ket{\psi_{\chi=2}^{(k)}}=\ket{0}
\end{equation}
are hold, $L[U]^{(k)\dagger}$ can be considered a disentangler, transforming the quantum state into a product state.
Finally, a new MPS $\ket{\psi^{(k+1)}}=\mathrm{L}[U]^{(k)\dagger}\ket{\psi^{(k)}}$ can be obtained.
Since $L[U]^{(k)\dagger}$ acts as a disentangler, $\ket{\psi^{(k+1)}}$ should have reduced entanglement compared to $\ket{\psi^{(k)}}$.
By repeating this process starting from the original MPS $\ket{\psi^{(0)}}$, a quantum circuit $\prod_{k=1}^K \mathrm{L}[U]^{(k)}\ket{0}$ with multiple layers can be generated.
 
% decomposition by optimization
Optimization-based methods are employed as an alternative approach to embedding tensor networks into quantum circuits.
This method sequentially optimizes the unitary operators within the quantum circuit to maximize the magnitude of the inner product between the quantum circuit and the tensor network.
Evenbly and Vidal proposed an iterative optimization technique that utilizes the calculation of environment tensors and SVD~\cite{env-tensor}.
Similarly, Shirakawa et al. employed this iterative optimization method to embed quantum states into quantum circuits~\cite{mpsoptim}.
An environment tensor is calculated to update a unitary by removing the unitary from the circuit and contracting the remaining tensor network.
We perform the SVD of the environment tensor and utilize the fact that the resulting unitary matrix serves as the optimal operator to increase the fidelity between the original tensor network and the constructed quantum circuit.
For a more detailed explanation of the optimization algorithm, please refer to \cite{mpsdecomp,mpsoptim} or the chapter on optimization algorithms for TTNs described later.

% decomposition of MPS
Rudolph et al. investigated the optimal sequence for combining these two methods to achieve the highest accuracy~\cite{mpsdecomp}.
The study concludes that the best results are obtained by adding new layers through a systematic decomposition step and optimizing all layers with each addition.
This method has achieved exceptional accuracy in embedding a wide range of MPSs in fields such as physics, machine learning, and random systems.
Therefore, it can be considered one of the best current techniques for embedding MPSs into shallow circuits.
However, this method does not support embedding tensor networks other than MPSs.
Consequently, embedding tensor networks such as TTNs and other non-MPS structures remains an unresolved issue.

\section{\uppercase{Proposed Method}}

We propose a method for embedding TTNs into shallow quantum circuits.
Any shape of TTN can be converted into a binary tree using SVD~\cite{structure-optim}, so we assume a binary tree for simplicity.
However, this method can be generalized to embed TTNs of any shape.
The fundamental approach of the proposed method is similar to that of MPSs~\cite{mpsdecomp}. 
By repeatedly adding layers using the systematic decomposition algorithm and optimizing the entire circuit, we generate highly accurate embedded quantum circuits.
This section first introduces the systematic decomposition algorithm for TTNs.
Due to the increased structural complexity of TTNs compared to MPSs, the contraction methods become non-trivial. 
We propose a method that balances approximation error and computational cost.
Next, we describe the optimization algorithm for TTNs and integrate it with the systematic decomposition.
Finally, we discuss the computational cost and demonstrate that embedding TTNs is feasible within practical computational limits.

\subsection{Systematic Decomposition}
Our systematic decomposition algorithm for TTNs is an extension of the algorithm for MPSs introduced in \cite{EncodingMPS}.
We denote the original TTNs as $\ket{\psi_0}$, the $k$-th layer of the quantum circuit as $\mathrm{L}[U]^{(k)}$, the number of layers in resulting quantum circuit as $K$, and the resulting quantum circuit as $\prod_{k=1}^K \mathrm{L}[U]^{(k)}\ket{0}$.
Algorithm~\ref{algorithm:decomposition} details the systematic decomposition process for TTNs, as depicted in Figure~\ref{fig:decomposition}(a).

\begin{algorithm}[!h]
 \caption{Systematic decomposition.}
 \label{algorithm:decomposition}
 \KwIn{TTN$\ket{\psi_0}$,Maximum layers $K$}
 \KwOut{Quantum Circuit $\prod_{k=1}^K \mathrm{L}[U]^{(k)}\ket{0}$}
 $\ket{\psi^{(1)}} \leftarrow{\ket{\psi_0}}$\;
 \For{$k=1$ to $K$}{
 Truncate $\ket{\psi^{(k)}}$ to $\ket{\psi^{(k)}_{\chi=2}}$ via SVD\;
 Convert $\ket{\psi^{(k)}_{\chi=2}}$ to $\mathrm{L}[U]^{(k)}$\;
 $\ket{\psi^{(k+1)}} \leftarrow{\mathrm{L}[U]^{(k)\dagger}}\ket{\psi^{(k)}}$\;
 }
\end{algorithm}

In this algorithm, a copy of the original TTN is truncated to a lower dimension using SVD.
Truncating bond dimension $\chi$ TTNs to bond dimension two TTNs can be done accurately, similar to MPSs, by shifting the canonical center so that the local SVD matches the global SVD.
This truncated TTN is then transformed into a single layer of two-qubit gates by converting the isometric tensors in the layer into unitary tensors using the Gram-Schmidt orthogonalization process.
 The inverse of this layer is applied to the original TTN, resulting in a partially disentangled state with potentially reduced entanglement and bond dimensions.
This process can be iteratively repeated to generate multiple layers, which are indexed in reverse order to form a circuit that approximates the target TTN. 
Notably, the final layer of the disentangling circuit is used as the initial layer of the quantum circuit for approximation.

While this algorithm appears to function effectively at first glance, the computation of $\ket{\psi^{(k+1)}} \leftarrow{\mathrm{L}[U]^{(k)\dagger}}\ket{\psi^{(k)}}$ is exceedingly challenging from the perspective of tensor networks.
The structure of $\ket{\psi^{(k+1)}}$ is TTN, whereas tensor network $\mathrm{L}[U]^{(k)\dagger}\ket{\psi^{(k)}}$ has a more complex structure, necessitating its transformation into the shape of TTN.
As previously mentioned, naively transforming a general tensor network can require exponentially large memory relative to the number of qubits.
Additionally, although limiting the bond dimension during the transformation can facilitate the process, the sequence of transformations can lead to substantial approximation errors.

To address the issue of tensor network transformations arising from the complexity of TTN structures, we employ a penetration algorithm as a submodule.
As illustrated in Figure~\ref{fig:decomposition}(b), the penetration algorithm operates by contracting two tensors connected by a single edge along the connected axis, appropriately reordering the axes, and then separating them using SVD.
This makes it appear as if the positions of the two tensors have swapped.
Additionally, by adjusting the number of singular values retained during the SVD, we can balance approximation accuracy and computational cost.

\begin{algorithm}[!h]
 \caption{Transformation of tensor networks using the penetration algorithm.}
 \label{algorithm:transformation}
 \KwIn{$\mathrm{L}[U]^{(k)\dagger}\ket{\psi^{(k)}}$}
 \KwOut{TTN $\ket{\psi^{(k+1)}}$}
 \For{$i=N-1$ to $1$}{
 Split $\mathrm{L}[U]^{(k)\dagger}_i$ into $U_p$ and $L_o$ via SVD\;
 \While{$U_p$ is not connected to $A^{(i)}$}{
 $A \leftarrow$  $U_p$'s left tensor\;
 Make $U_p$ penetrate $A$\;
 }
 \While{$L_o$ is not conntected to $A^{(i)}$}{
 $A \leftarrow$  $L_o$'s left tensor\;
 Make $L_o$ penetrate $A$\;
 }
 $A^{(i)} \leftarrow$ Contract $A^{(i)}$, $U_p$, and $L_o$  \;
 }
\end{algorithm}

Algorithm~\ref{algorithm:transformation} details the transformation of tensor networks using the penetration algorithm, as depicted in Figure~\ref{fig:decomposition}(c).
We assign numbers to $\ket{\psi^{(k)}}$ in breadth-first search~(BFS) order from the root.
We denote $i$-th tensor of $\ket{\psi^{(k)}}$ as $A^{(i)}$.
We also assign numbers to $\mathrm{L}[U]^{(k)\dagger}$ based on its origin in the TTN, denoting the tensor with number $i$ as $\mathrm{L}[U]^{(k)\dagger}_i$.
For each tensor $\mathrm{L}[U]^{(k)\dagger}_i$ in $\mathrm{L}[U]^{(k)\dagger}$, first use SVD to split it into upper tensor $U_p$ and lower tensor $Lo$ to reduce the computational complexity in the penetration algorithm.
Then apply the penetration algorithm iteratively until the upper tensor is connected with $A^{(i)}$.
Repeat this for the lower tensor too.
Finally, contract the upper and lower tensors with $A^{(i)}$ to form a new TTN's $i$-th tensor.
Perform this process sequentially from the highest-numbered tensor in $\mathrm{L}[U]^{(k)\dagger}$.
This algorithm allows contraction with each tensor in $\mathrm{L}[U]^{(k)\dagger}$ without changing the TTN structure of $\ket{\psi^{(k)}}$.
Consequently, despite minor approximation errors during penetration, $\mathrm{L}[U]^{(k)\dagger}\ket{\psi^{(k)}}$ can be transformed into a TTN structure.

\begin{figure*}[!h]
  \centering
  \includegraphics[width=\linewidth]{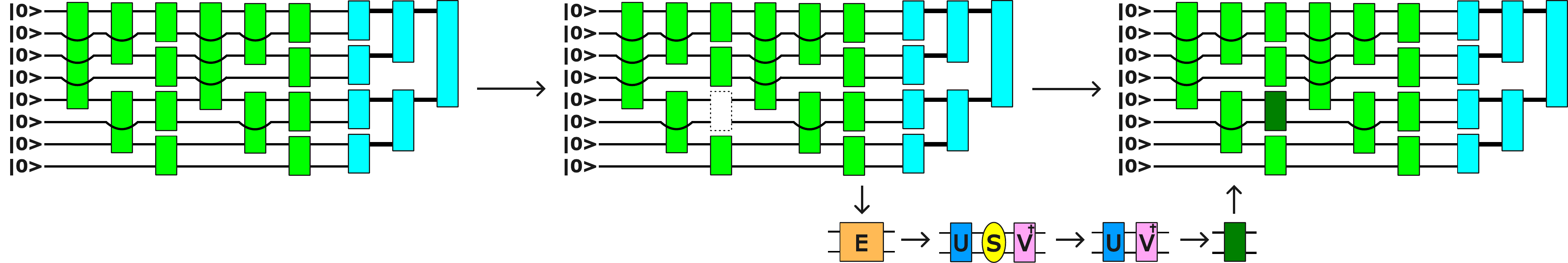}
   % {\epsfig{file = fig-optimization.pdf, width = 15cm}}
  \caption{The environment tensor $E$ is obtained by contracting all tensors except the one of interest. We compute its SVD $E=USV^\dagger$ and calculate $UV^\dagger$ to find the unitary operator closest to the environment tensor. The generated unitary operator is positioned at the location of the removed unitary operator.}
  \label{fig:optimization}
 \end{figure*}

\subsection{Integration with Optimization}
As demonstrated in \cite{mpsdecomp}, the systematic decomposition algorithm achieves the highest embedding accuracy when appropriately combined with the optimization algorithm.
In this study, we also integrate systematic decomposition with the optimization algorithm.
The fundamental concept of the optimization algorithm for quantum circuits embedded with TTNs is analogous to that for quantum circuits embedded with MPS. 
The primary difference lies in the ansatz of the quantum circuits; however, the optimization algorithm can be executed in a similar manner.

\begin{algorithm}[!h]
 \caption{Optimization.}
 \label{algorithm:optimization}
 \KwIn{Quantum Circuit $\prod_{k=1}^K \mathrm{L}[U]^{(k)}\ket{0}$, number of sweeps $T$, learning rate $r \in [0,1]$}
 \KwOut{Optimized Quantum Circuit $\prod_{k=1}^K \mathrm{L}[U]^{(k)}\ket{0}$}
 \For{$t=1$ to $T$}{
  \For{$k=1$ to $K$}{
   \For{$i=1$ to $N-1$}{
    $U_{old} \leftarrow \mathrm{L}[U]^{(k)}_i$\;
    Calculate environment tensor $E$\;
    SVD $E=USV^\dagger$\;
    $U_{new} \leftarrow UV^\dagger$\;
    $\mathrm{L}[U]^{(k)}_i \leftarrow U_{old}(U_{old}^\dagger U_{new})^r$\;
   }
  }
 }
\end{algorithm}

Algorithm~\ref{algorithm:optimization} details the optimization process for TTNs, as depicted in Figure~\ref{fig:optimization}.
The environment tensor $E$ is obtained by contracting all tensors except for the one of interest.
In this algorithm, the tensor of interest is $\mathrm{L}[U]^{(k)}_i$ and $E$ becomes a four-dimensional tensor with two legs on the left and two on the right.
We compute the SVD of $E$ and utilize the fact that the product $UV^\dagger$ is the unitary matrix that maximizes the magnitude of the inner product between the original TTN and the generated quantum circuit~\cite{mpsoptim}.
Given the strength of this local update, we introduce a learning rate $r$, which modifies the unitary update rule via $U_{old}(U_{old}^\dagger U_{new})^r$. 
Replacing $\mathrm{L}[U]^{(k)}_i$ with the unitary operator calculated in this manner for all operators constitutes one step, and repeating this process for $T$ steps completes the optimization algorithm.

The integration of the systematic decomposition and optimization algorithms involves a method where a new layer is added using the systematic decomposition algorithm, followed by optimizing the entire quantum circuit with the optimization algorithm.
This process is repeated iteratively.
Rudolph et al. refer to this method as Iter[$D_i$, $O_{all}$], confirming that it achieves the highest accuracy regardless of the type of MPSs~\cite{mpsdecomp}.
When creating the $k+1$-th layer using the systematic decomposition algorithm, the layers up to $k$ are first absorbed into the original TTN using the penetration algorithm before executing the systematic decomposition algorithm.
It should be noted that since we are using the penetration algorithm, the $\ket{\psi}$ that has absorbed $L[U]^{j\dagger}$ retains its TTN structure.
The integrated algorithm is presented as Algorithm~\ref{algorithm:integration}.
\begin{algorithm}[!h]
 \caption{Proposed method.}
 \label{algorithm:integration}
 \KwIn{TTN$\psi_0$,Maximum layers $K$}
 \KwOut{Quantum Circuit $\prod_{k=1}^K \mathrm{L}[U]^{(k)}\ket{0}$}
 $\ket{\psi} \leftarrow{\ket{\psi_0}}$\;
 \For{$k=1$ to $K$}{
 Truncate $\ket{\psi}$ to $\ket{\psi_{\chi=2}}$ via SVD\;
 Convert $\ket{\psi_{\chi=2}}$ to $\mathrm{L}[U]^{(k)}$\;
 Optimize $\prod_{k'=1}^{k} \mathrm{L}[U]^{(k')}$\;
 \For{$j=1$ to $k$}{
 Absorb $\mathrm{L}[U]^{j\dagger}$ into $\ket{\psi}$\;
 }
 }
\end{algorithm}

\subsection{Computational Complexity}
For a TTN with a maximal bond dimension $\chi$, the memory requirements scale $\mathcal{O}(N\chi^3)$.
However, the computational complexity of transforming the TTN into its canonical form scale $\mathcal{O}(N\chi^4)$.
Most of the algorithms related to TTNs require transformation into canonical form.
Therefore, it is reasonable to assume that we set $\chi$ such that computations of $\mathcal{O}(N\chi^4)$ can be performed efficiently.

In Algorithm~\ref{algorithm:integration}, truncating $\ket{\psi}$ to $\ket{\psi_{\chi=2}}$ requires $\mathcal{O}(N\chi^3)$.
Optimizing the entire circuit is significantly more challenging compared to MPS due to the complexity of the TTN structure.
Generally, exponential memory is required with respect to $N$, but by using an appropriate contraction order and caching, the computation can be performed in $\mathcal{O}(N\chi^3 4^K)$, which is linear in $N$.
Although the computational complexity increases exponentially with the number of layers, this algorithm is designed for embedding into shallow quantum circuits, and it operates efficiently for $K \approx 7$, as used in our experiments.
Furthermore, when embedding TTNs into a larger number of layers, it is possible to achieve $\mathcal{O}(N\log{N}\chi^4KT)$, where $T$ is the number of sweeps, linear computational complexity with respect to the number of layers, bond dimension by introducing approximations that prevent the bond dimension from exceeding $\chi$ during the contraction of environment tensors, as used in the embedding of MPS~\cite{mpsdecomp}.

The computational complexity of absorbing $K$ layers into the original TTN is $\mathcal{O}(N\log{N}\chi^4K)$.
In the penetration algorithm, the SVD of a matrix with row dimension $\chi^2$ and column dimension $\chi$ has a complexity of $\mathcal{O}(\chi^4)$.
Given that the depth of the TTN is $\mathcal{O}(\log{N})$, the penetration operation is performed $\mathcal{O}(\log{N})$ times per tensor.
Since each layer contains $\mathcal{O}(N)$ tensors, the complexity of absorbing one layer is $\mathcal{O}(N\log{N}\chi^4)$, leading to a total complexity of $\mathcal{O}(N\log{N}\chi^4K)$ for $K$ layers.
In summary, the computational complexity to generate one layer is $\mathcal{O}(\max (N\log{N}\chi^4 K, N\chi^3 4^K))$, and it is $\mathcal{O}(\max (N\log{N}\chi^4 K^2, N\chi^3 4^K))$ for $K$ layers.
It scales with the number of qubits.
Additionally, the bond dimension is constrained to $\chi^4$, making it suitable for embedding TTNs with large bond dimensions.
Furthermore, by allowing approximations in optimization, the complexity can be reduced to $\mathcal{O}(N\log{N}\chi^4 K^2T)$ for $K$ layers.

\begin{figure*}[!h]
  \centering
   % {\epsfig{file = fig_ttn_mps.pdf, width = 11.5cm}}
   \includegraphics[width=0.75\linewidth]{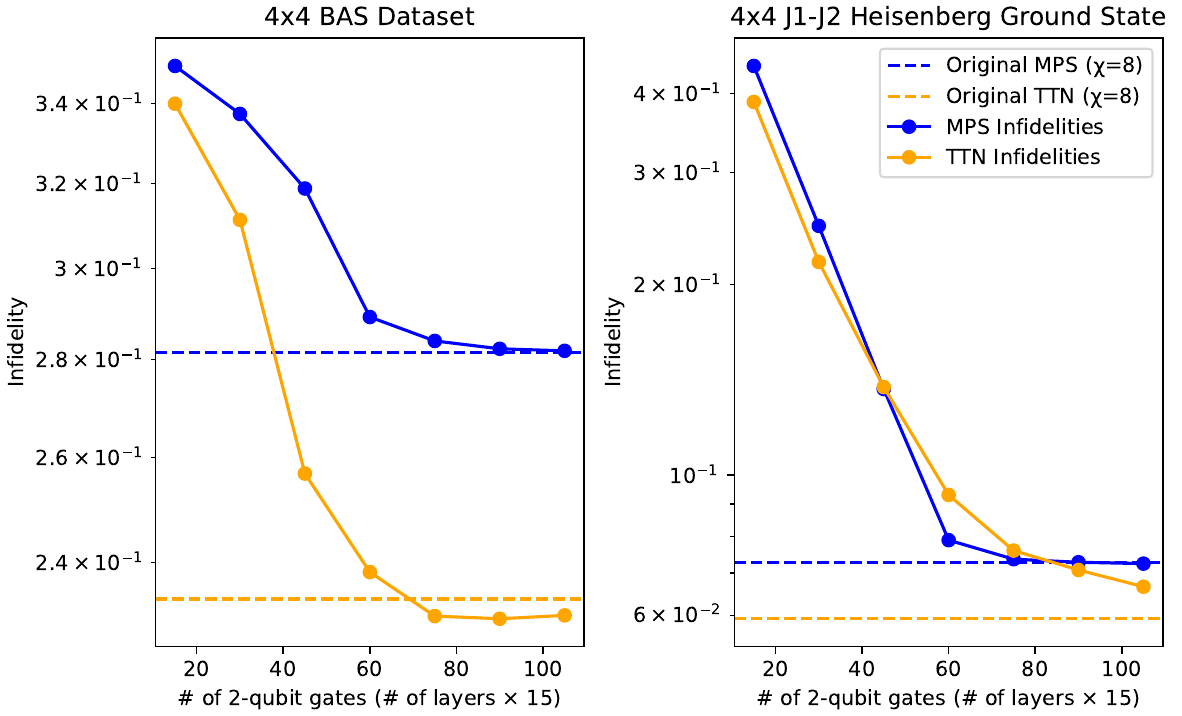}
  \caption{Infidelities between original state vectors and generated quantum circuits. The bond dimensions of tensor networks are eight and the number of optimization steps is 1000. The learning rate for optimization is set to 0.65 for the BAS Dataset and 0.6 for the $J_1$-$J_2$ Heisenberg model. MPS is represented by the blue line, while TTN is depicted by the orange line. Additionally, the infidelity between the tensor network and the state vector is indicated by the dotted line.}
  \label{fig:result1}
 \end{figure*}

 \begin{figure*}[!h]
  \centering
   % {\epsfig{file = fig_optim_type, width = 11.5cm}}
   \includegraphics[width=0.75\linewidth]{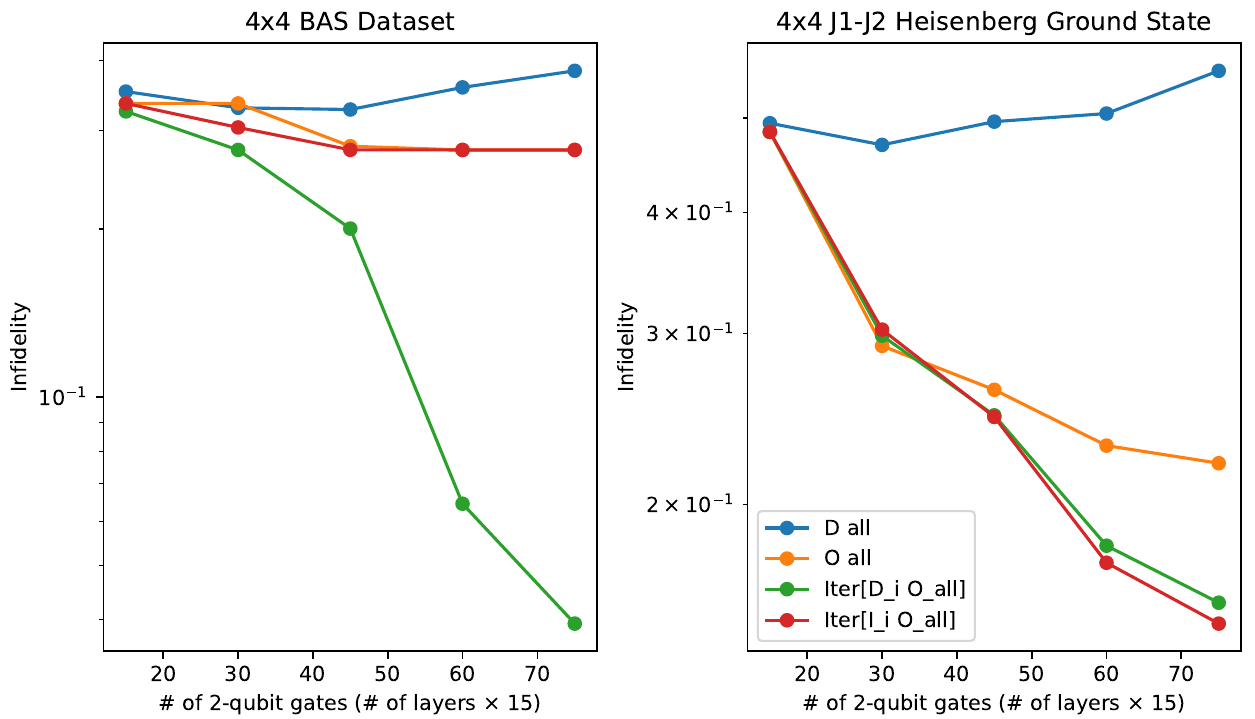}
  \caption{Infidelities between original TTNs and generated quantum circuits. The bond dimensions of TTNs are 16 and the number of optimization steps is 1000. The learning rate for optimization is set to 0.6.}
  \label{fig:result2}
 \end{figure*}
 
\section{\uppercase{Experiments}}
We conducted experiments using two distinct state vectors from the fields of machine learning and physics.
The first state vector represents a uniform superposition over the binary data samples in the $4\times4$ bars and stripes~(BAS) dataset~\cite{bas}, which has become a canonical benchmark for generative modeling tasks in quantum machine learning.
The second state vector represents the ground state of the $J_1$-$J_2$ Heisenberg model, a model that characterizes competing interactions in quantum spin systems.
The Hamiltonian for this model is given by the following equation,
\begin{equation}
    H=J_1 \sum_{<i,j>} \bm{S}_i \cdot \bm{S}_j + J_2 \sum_{<<i,j>>} \bm{S}_i \cdot \bm{S}_j,
\end{equation}
where $J_1$ ($J_2$) represents the nearest~(next-nearest) neighbor interactions. 
By varying the ratio of the first and second nearest-neighbor interactions, the $J_1$-$J_2$ Heisenberg model generates complex quantum many-body phenomena and has been widely studied.
In this paper, we utilized $J_2/J_1=0.5$.
Both state vectors are two-dimensional systems with long-range correlations, making them more suitably represented by TTNs rather than MPSs.

The state vector of the BAS was manually prepared, while the ground state of the $J_1$-$J_2$ Heisenberg model was generated using a numerical solver package $\mathcal{H}\Phi$~\cite{hphi}, which is designed for a wide range of quantum lattice models. 
MPSs and TTNs were constructed by iteratively applying SVD to the state vector from the edges.

The conversion from MPSs and TTNs to quantum circuits employed the Iter[$D_i$, $O_{all}$] method, which is also utilized in the proposed method.
Additionally, to compare with the proposed method, we also used $D_{all}$, $O_{all}$, and Iter[$I_i$, $O_{all}$] from \cite{mpsdecomp}.
The $D_{all}$ method generates circuits solely through systematic decomposition.
The $O_{all}$ method optimizes circuits starting from an initial state composed only of identity gates.
The Iter[$I_i$, $O_{all}$] method sequentially adds identity layers, optimizing the entire circuit at each step.
The number of sweeps in the optimization process was set to 1000.
Experiments were conducted using various learning rates ranging from 0.5 to 0.7, and the rate that demonstrated the highest convergence accuracy was selected.

To measure the accuracy of embedding into quantum circuits, we utilized infidelity as the evaluation metric.
The infidelity $I_f$ between two quantum states, $\ket{\Psi}$ and $\ket{\Phi}$, is expressed by
\begin{equation}
    I_f = 1 - |\braket{\Phi|\Psi}|,
\end{equation}
and a smaller infidelity indicates that the two quantum states are closer.
In this study, infidelity quantifies the success of our transformations.

Figure~\ref{fig:result1} illustrates the infidelity between the original state vector and the quantum circuits embedded with either TTN or MPS.
Despite the greater difficulty in embedding TTNs into shallow quantum circuits due to their hierarchical structure, the infidelities between original state vectors and generated quantum circuits from TTNs are sufficiently low, indicating the successful embedding of the TTNs into shallow quantum circuits.
Notably, in the BAS dataset, the TTN's superior representational capacity enables more accurate embeddings than MPS.
The result of the $J_1$-$J_2$ Heisenberg model reveals an intriguing pattern: TTN outperforms in both shallow and deep quantum circuits, while MPS excels in the intermediate region.
The superior performance of TTN in shallow circuits is due to the minimal loss from the penetration algorithm, enabling high-precision systematic decomposition.
In deep circuits, TTN's higher representational capacity leads to better convergence accuracy, as indicated by the dotted lines.
Thus, the $J_1$-$J_2$ model results enhance our understanding of the differences between MPS and TTN embeddings.

Figure~\ref{fig:result2} highlights the importance of the systematic decomposition algorithm for embedding TTNs into quantum circuits.
As demonstrated in previous research on MPS~\cite{mpsdecomp}, methods such as Iter[$D_i$, $O_{all}$] and Iter[$I_i$, $O_{all}$], which sequentially add layers and optimize all circuits, achieve high accuracy. 
Furthermore, as suggested in previous studies, it was confirmed that Iter[$I_i$, $O_{all}$] struggles with the BAS dataset even in TTN embeddings, where the singular values at the bond become discontinuous. 
This indicates that Iter[$D_i$, $O_{all}$] is the most effective embedding method, achieving high accuracy in multiple fields, including machine learning and physics, thus affirming the critical importance of the systematic decomposition algorithm.
This result also verifies the effective performance of the proposed systematic decomposition method.

\section{\uppercase{Conclusion}}
To avoid the barren plateau issue in VQAs, there is increasing interest in using tensor networks to initialize quantum circuits. 
However, embedding tensor networks into shallow quantum circuits is generally difficult, and prior research has been limited to embedding MPSs.
In this study, we propose a method for embedding TTNs, which have a more complex structure than MPSs and can efficiently represent two-dimensional systems and systems with long-range correlations, into shallow quantum circuits composed solely of two-qubit gates.
We applied our proposed method to various types of TTNs and confirmed that it prepares quantum circuit parameters with better accuracy than embedding MPSs.
Additionally, the computational complexity is $\mathcal{O}(\max (N\log{N}\chi^4 K^2, N\chi^3 4^K))$, or $\mathcal{O}(N\log{N}\chi^4 K^2T)$ with approximation, making it applicable to practical problems.
This study will serve as an important bridge for implementing hybrid algorithms combining tree tensor networks and quantum computing.

\section*{\uppercase{Acknowledgements}}
This work was supported by the Center of Innovation for Sustainable Quantum AI, JST Grant Number JPMJPF2221, and by Japan Society for the Promotion of Science KAKENHI, Grant Numbers 22K18682 and 23H03818.
We acknowledge the use of Copilot~(Microsoft, \url{https://copilot.microsoft.com/}) for the translation and proofreading of certain sentences within our paper.

\bibliographystyle{apalike}
{\small
\bibliography{main}}

% \section*{\uppercase{Appendix}}

\end{document}